\documentclass[conference]{IEEEtran}
\IEEEoverridecommandlockouts
\usepackage{cite}
\usepackage{amsmath,amssymb,amsfonts}
\usepackage{graphicx}
\usepackage{xcolor}
\usepackage{caption}
\usepackage{adjustbox}
\usepackage{tabularray}
\usepackage{url}
\usepackage{multirow}

\def\BibTeX{{\rm B\kern-.05em{\sc i\kern-.025em b}\kern-.08em
    T\kern-.1667em\lower.7ex\hbox{E}\kern-.125emX}}

\makeatletter
\newcommand{\newlineauthors}{%
  \end{@IEEEauthorhalign}\hfill\mbox{}\par
  \mbox{}\hfill\begin{@IEEEauthorhalign}
}
\makeatother

\title{Speech-Based Estimation of Schizophrenia Severity Using Feature Fusion}

\author{\IEEEauthorblockN{Gowtham Premananth and  Carol Espy-Wilson\thanks{This work was supported by the National Science Foundation grant numbered 2124270.}}
\IEEEauthorblockA{
\textit{Dept. of Electrical and Computer Engineering, University of Maryland College Park, MD, USA},
\\
gowtham8@umd.edu,
espy@umd.edu\vspace{-3em}}

}

\begin{document}
\maketitle
\vspace{-12pt}
\begin{abstract}
Speech-based assessment of the schizophrenia spectrum has been widely researched over in the recent past. In this study, we develop a deep learning framework to estimate schizophrenia severity scores from speech using a feature fusion approach that fuses articulatory features with different self-supervised speech features extracted from pre-trained audio models. We also propose an auto-encoder-based self-supervised representation learning framework to extract compact articulatory embeddings from speech. Our top-performing speech-based fusion model with Multi-Head Attention (MHA) reduces Mean Absolute Error (MAE) by 9.18\% and Root Mean Squared Error (RMSE) by 9.36\% for schizophrenia severity estimation when compared with the previous models that combined speech and video inputs.
\end{abstract}

\begin{IEEEkeywords}
Schizophrenia, Self-supervised speech representations, Articulatory coordination features, Feature fusion
\end{IEEEkeywords}
\vspace*{-4pt}
\section{\textbf{Introduction}}
\vspace*{-3pt}
Schizophrenia is a chronic and severe mental health disorder that affects approximately 24 million people globally \cite{b1}. Schizophrenia affects how individuals interpret reality, often leading to a fragmented perception of the world around them. It is characterized by a broad spectrum of symptoms varying in severity, which disrupt perceptions of reality, cause notable behavioral changes, and impair daily functioning. Some of the common symptoms of schizophrenia include hallucinations, disorganized thinking, reduced emotional expression, and poverty of speech. This diversity in symptoms, especially regarding speech, has spurred interest in using speech as a biomarker for detecting and evaluating schizophrenia. Clinicians use assessment scales like the Brief Psychiatric Rating Scale (BPRS) \cite{b2} to measure psychiatric symptoms presented by subjects based on the severity of the symptom.

The initial works focusing on using speech as a biomarker for different mental health disorders used inputs like statistical acoustical features \cite{b3}, Mel-Frequency Cepstral Coefficients (MFCCs) \cite{b4}, and spectrograms \cite{b5}. With the recent development of self-supervised representation learning models, different pre-trained speech representation models like Wav2Vec2\cite{b6}, and  WavLM\cite{b7} have been introduced. As these self-supervised speech representations were trained on large corpora comprising hours of speech recordings from different speakers in different environments, these representations are more generalizable and therefore have been effectively used in different mental health assessment systems \cite{b8,b9}. Apart from these typical acoustic features of speech, our research has also focused on utilizing articulatory features to assess different mental health disorders \cite{b10,b11}.

Deep learning-based multimodal fusion approaches have been heavily researched recently and have produced promising results in different domains where input data from various modalities are used to outperform conventional unimodal systems \cite{b12,b24}. These fusion models have proven effective in data-scarce settings where training data from a single modality is insufficient to train deep learning models\cite{b23}. Fusion models utilize input from multiple modalities and extract cross-modal information and complementary information from different modalities to compensate for the lack of training samples. Because of these characteristics, multimodal fusion approaches have been a perfect fit for assessing various mental health disorders given datasets used in medical applications are comparatively small. Recent works on multimodal assessment of disorders like depression \cite{b10} and schizophrenia \cite{b11,b13,b14} have shown improved performance by fusing inputs from multiple modalities. 

The fusion approaches and models mentioned above have been mainly used on data from different modalities. The availability of data from multiple modalities, especially in medical domain applications, is not always feasible due to privacy concerns and other legal limitations. Therefore these fusion approaches have to be used in a way to get the most out of the available modalities. For this reason, we utilize feature fusion to fuse information from two largely independent feature spaces extracted from speech.

The main contributions of the paper are as follows:
\begin{enumerate}
    \item A self-supervised articulatory representation learning model to produce concise articulatory representations from vocal tract variables (TVs).
    \item A feature fusion model for speech-based schizophrenia severity score estimation that utilizes self-supervised representations and concise articulatory representations.
\end{enumerate}

\section{\textbf{Dataset}}
\vspace*{-3pt}

\begin{figure}[h!]
    \centering
    \includegraphics[width=80mm, height= 60mm]{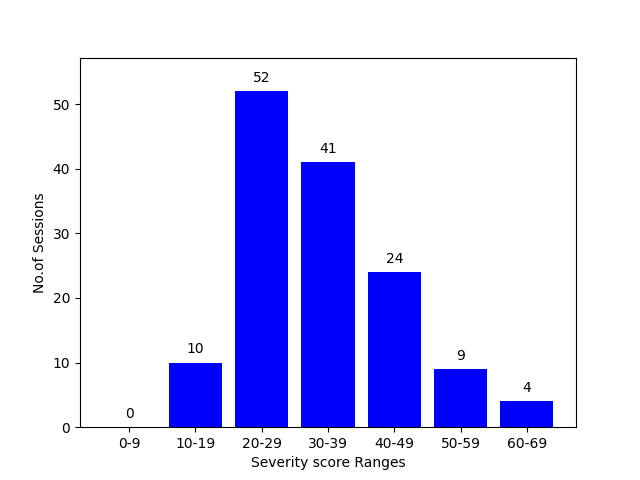}
    \caption{\textbf{Distribution of the severity scores in the dataset}}
    \label{fig:sessions} 
\end{figure}

The dataset used in this study was collected as a part of an interdisciplinary mental health assessment project \cite{b15} jointly carried out by the University of Maryland School of Medicine and the University of Maryland - College Park. The dataset contains subjects with schizophrenia and depression with varying severity of symptoms and healthy controls. The dataset comprises audio and video recordings along with the transcripts of the interview sessions between the subjects and an investigator. As this study only focuses on the audio modality, 140 audio recordings that belong to 40 subjects (healthy controls and schizophrenia subjects) who have attended multiple sessions over six weeks are used in this study. The investigators assessed the subjects using the BPRS scale before each session. The BPRS scale contains 18 items each corresponding to a symptom. Each item is given a score in the range of 1-7 depending on the severity of the symptom: the higher the score the more severe the symptom. The total BPRS score is used as the severity score in this study. The distribution of the severity scores associated with sessions used in this study is shown in Fig \ref{fig:sessions}.

\vspace*{-14pt}
\vspace*{-12pt}

\section{\textbf{Data Preprocessing and Feature Extraction}}
\vspace*{-3pt}

The audio recordings in the dataset contain the speech of both the subject and the investigator. Initially, the audio recordings were diarized according to their transcripts to obtain the portions of the recordings where the subject was speaking. After diarization, the audio was segmented into 40-second segments. The segmented audio was then used in all different feature extraction frameworks to obtain different self-supervised speech representations and articulatory features.

\subsection{\textbf{Articulatory features}}
\vspace*{-3pt}
For articulatory features, a combination of 6 vocal Tract Variables (TVs), and 2 glottal parameters are used. An acoustic-to-articulatory speech inversion system \cite{b16} is used to estimate the 6 TVs related to the constriction degree and location of the articulators (lips, tongue body, and tongue tip). The glottal parameters aperiodicity and periodicity are estimated using an APP (Aperioidicty Periodicity Pitch) detector \cite{b17}. To get at the phasing between articulatory gestures, we convert the TVs into TV-based Full Vocal Tract Coordination (FVTC) features (high-level correlation matrix structures). The TV-based FVTC features are constructed based on the channel-delay correlation of the features \cite{b18}. The delayed correlation between $TV_1$ and $TV_2$ delayed by $d$ frames $( r_{TV_1 , TV_2}^d)$, is computed using Eq.\ref{eq:delay} where $N$ is the total number of frames in the input.
\vspace*{-3pt}
\begin{equation}
\label{eq:delay}
r_{TV_1 , TV_2}^d=\frac{\sum_{t=0}^{N-d-1}TV_1[t]TV_2[t+d]}{N-|d|}
\end{equation}

The correlation vector for the pair of TVs ($TV_1$ and $TV_2$) is constructed by stacking the delayed correlations between them with delays $d \in [0, D]$ as shown in Eq. \ref{eq:vector}.  $D$ here is purely a design choice.
\vspace*{-3pt}
\begin{equation}
\label{eq:vector}
R_{TV_1 , TV_2}=[r_{TV_1 , TV_2}^0 , r_{TV_1 , TV_2}^1 , ..... r_{TV_1 , TV_2}^D]^T
\end{equation}

TV-based FVTC features are finally constructed by stacking auto-correlation and cross-correlation vectors generated for all pairs of TVs $[(TV_1,TV_1),(TV_1,TV_2),...,(TV_1,TV_8),....,(TV_8,TV_8)]$. as shown in Eq. \ref{eq:fvtc} 
\vspace*{-3pt}
\begin{equation}
\label{eq:fvtc}
R_{FVTC}=[R_{TV_1 , TV_1} , ... ,R_{TV_1 , TV_8}, .. ,R_{TV_8 , TV_8}]^T
\end{equation}

\subsection{\textbf{Self-supervised speech representations}}
\vspace*{-3pt}
For each audio segment, self-supervised representations were extracted from their respective pre-trained models (Wav2vec2.0 \cite{b6} and WavLM\cite{b7}). As different pre-trained self-supervised models use different-sized contextualized windows in the audio segment to extract representations, the representation for the segment was obtained by taking the mean of the representations extracted for all the contextual windows in the segment. Wav2Vec2.0-base and WavLM-base-plus models produce a representation of size 768 while Wav2Vec2.0-large and WavLM-large models produce a representation of size 1024.

\vspace*{-8pt}
\begin{figure}[h!]
    \centering
    \includegraphics[width=85mm, height= 35mm]{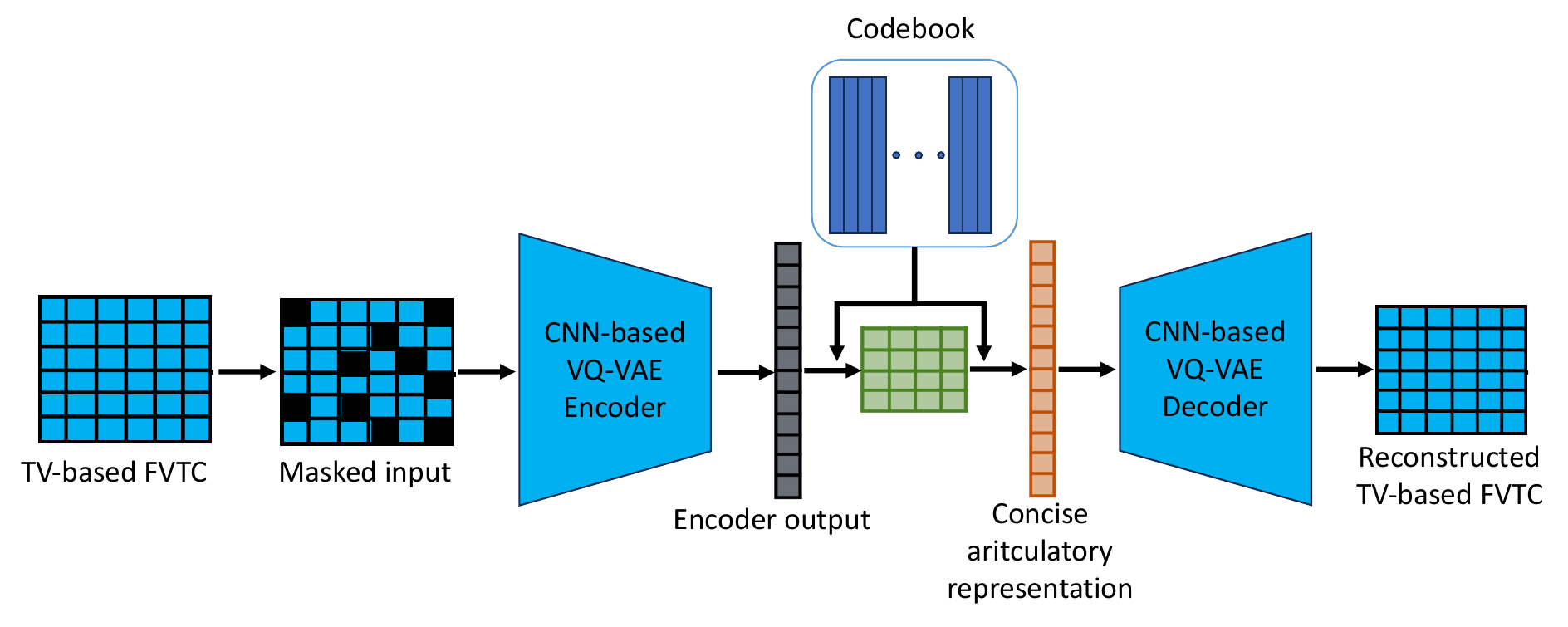}
    \caption{\textbf{VQ-VAE-based articulatory representation learning model}}
    \label{fig:articulatory_RL} 
\end{figure}
\vspace*{-8pt}

\section{\textbf{Model Architectures}}
\vspace*{-3pt}
\begin{figure*}[th!]
    \centering
    \includegraphics[width=170mm, height= 45mm]{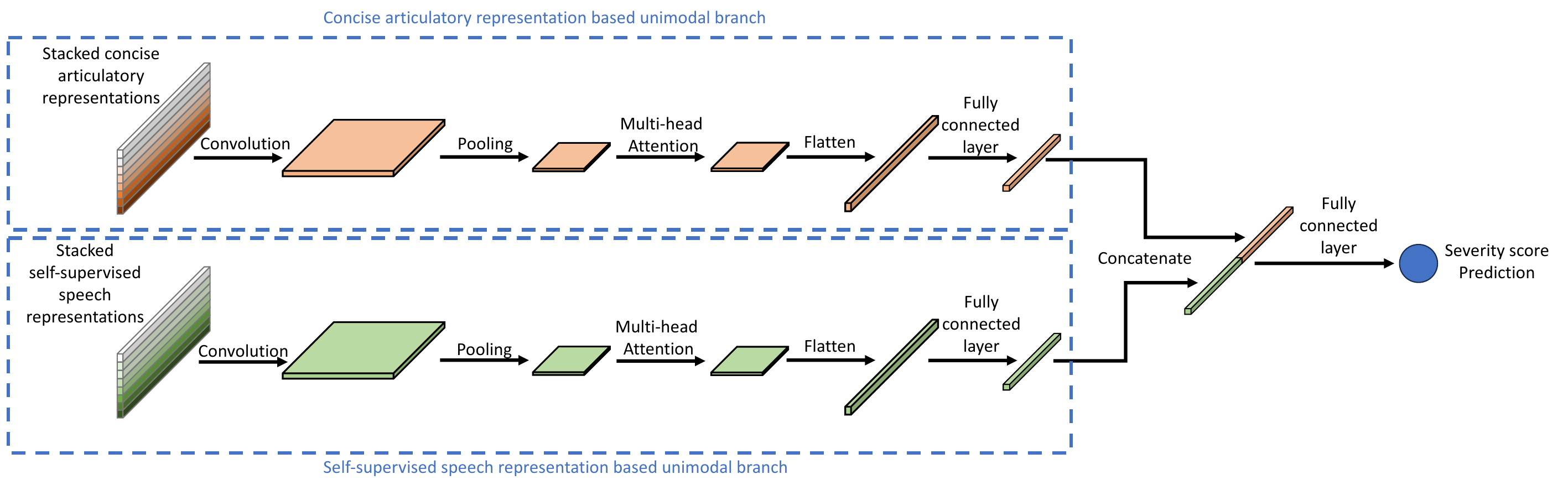}
    \caption{\textbf{Feature-fusion model architecture with concise articulatory representation and self-supervised representation based branches}}
    \label{fig:mrl}
\end{figure*}

In this work, we propose two model architectures. 
\begin{itemize}
    \item A self-supervised articulatory representation learning model to produce concise latent space representations from TV-based FVTC features.
    \item A feature fusion model for speech-based schizophrenia severity score estimation using self-supervised speech representations and self-supervised articulatory representation produced by the model mentioned above.
\end{itemize} 

\subsection{\textbf{Self-supervised articulatory representation learning model}}
\vspace*{-3pt}

Even though we have already converted TVs into TV-based FVTC features, those high-level correlation matrix structures contain a collection of auto-correlations and cross-correlations that still contain a certain degree of information redundancy in the feature space. To alleviate this redundancy issue and to create more concise articulatory representations, we propose a Vector Quantized Variational Auto Encoder \cite{b19} (VQ-VAE)-based representation learning model as shown in Fig.\ref{fig:articulatory_RL}. The VQ-VAE-based representation learning model takes TV-based FVTC feature matrices of each 40-second audio segment as input and tries reconstructing the input itself. For training purposes, a masked autoencoder-based approach was used where the input matrix was fed into the model after randomly masking out a pre-defined percentage of the input matrix. The encoder output quantized by the codebook is extracted as concise articulatory representations. The concise articulatory representations have an embedding size of 1024, which is a design choice made to keep it consistent with the embedding size of commonly used speech representations.

\subsection{\textbf{Feature fusion model architecture for speech-based schizophrenia severity score estimation}}
\vspace*{-3pt}
The feature fusion model as shown in Fig.\ref{fig:mrl} is constructed as two separate branches of CNN models that process the self-supervised speech representations and concise articulatory representations separately before passing them through Multi-Head Attention (MHA) \cite{b20} layers. After the attention layers, the attention outputs are fused by a simple concatenation layer and then sent through a fully connected layer to produce the estimated schizophrenia severity score.

\section{\textbf{Experiments}}
\vspace*{-3pt}

The dataset was divided into train, test, and validation folds with a ratio of 70:15:15 for all model training and evaluation purposes. The splits were divided in a subject-independent manner, where none of the segments nor the sessions belonging to a particular subject were allocated to two different folds. Initially, the self-supervised articulatory representation learning models were trained and evaluated. The representation learning models are trained based on total loss $(L_{Total})$ which consists of reconstruction loss $(L_{Reconstruction})$, commitment loss $(L_{Commitment})$, and codebook loss $(L_{Codebook})$. The total loss is calculated as follows
\vspace*{-3pt}
\begin{equation}
    L_{Total} = L_{Reconstruction} + \beta * L_{Commitment} + L_{Codebook}
\end{equation}

$\beta$ is a hyper-parameter that controls how much the encoder outputs commit to an embedding in the codebook. Along with $\beta$, the embedding size, the codebook latent dimension size, and the masking percentage of the input were tuned during training to find the best-performing representation learning model. The model was evaluated on the $Loss_{Total}$ for the test fold. All the self-supervised articulatory representation learning models were trained for 100 epochs with an initial learning rate of 5e-5 which was adjusted based on a reduce-on-plateau learning rate scheduler that had a patience of 25 epochs. 

The best-performing articulatory representation learning model was then used to extract the concise articulatory representations for all the audio segments. Likewise, speech representations were extracted using various pre-trained models. Then the extracted embeddings belonging to the same session were stacked together to create the session-wise embeddings. The session-wise embeddings were then used as inputs to train, test, and validate the feature fusion model. Hyper-parameters of the models were optimized using a manual grid search. The models were trained with different combinations of the number of convolution layers, the number of output channels in the convolution layers, dropout probabilities, the number of heads in multi-head attention layers, the number of fully-connected layers after fusion, and the number of neurons in the final fully-connected layer. 

\begin{table*}[th!]
  \caption{\centering
  \textbf{Performance comparison of schizophrenia severity estimation models}}
  \label{tab:results}
  \centering
  \begin{tabular}{ |l |l|l|c |c|c|}
  \hline
    \textbf{Model}& \textbf{Modalities}&\textbf{Features}&  \textbf{MAE}$\downarrow$ &\textbf{RMSE}$\downarrow$&\textbf{$\rho$}$\uparrow$\\
    \hline
    \hline
    Unimodal CNN model&Audio&Wav2vec2.0 - base&9.77&13.31&0.7451\\
    \hline
    Unimodal CNN model&Audio&Wav2vec2.0 - large&10.13&13.18&0.7532\\
    \hline
    Unimodal CNN model &Audio&WavLM - base-plus&10.17&14.09&0.7406\\
    \hline
    Unimodal CNN model&Audio&WavLM - large &9.84&13.67&0.5528\\
    \hline
    \hline
    Unimodal CNN model&Audio&TV-FVTC&8.89&10.32&0.6214\\
    \hline
    Unimodal CNN model&Audio&Concise articulatory representation&8.43&10.69&0.6275\\
    \hline
    \hline
    \textbf{Feature-Fusion with MHA} & \textbf{Audio}& \textbf{Concise articulatory representation, Wav2vec2.0 - base} & \textbf{6.53}& \textbf{8.04}&0.7587\\
    \hline
    Feature-Fusion with MHA&Audio&Concise articulatory representation, Wav2vec2.0 - large & 6.93&8.53&0.7443\\
    \hline
    Feature-Fusion with MHA&Audio&Concise articulatory representation, WavLM - base-plus & 7.82&9.74&0.7240\\
    \hline
    Feature-Fusion with MHA&Audio&Concise articulatory representation, WavLM - large &7.61&10.46&0.7089\\
    \hline
    \hline
    Multimodal \cite{b21}&Audio,video&TV-FVTC, FAU-FVTC&8.81&10.66&0.5754\\
    \hline
    Multimodal with multi-task learning \cite{b21}&Audio,video&TV-FVTC, FAU-FVTC&7.19&8.87&\textbf{0.7926}\\
\hline
  \end{tabular}
\end{table*}

The schizophrenia severity estimation task was only addressed in one of the previous works \cite{b21} through a multimodal (audio, video) representation learning framework that is used to extract task-agnostic multimodal representations for estimation. Therefore, as speech-based unimodal models were not trained for this specific task on this specific dataset, baseline unimodal models were also trained and evaluated for comparison purposes. Two separate unimodal models for self-supervised speech and concise articulatory representations were trained by replicating the same model architecture of the corresponding branch in the feature fusion model and a final fully connected layer for severity score estimation.

All the schizophrenia severity estimation models were trained for 400 epochs with an initial learning rate of 5e-4 which was adjusted based on a reduce-on-plateau learning rate scheduler that had a patience of 100 epochs. The models were optimized using an Adam optimizer for the mean squared error loss. The trained models were evaluated based on the test set's Mean Absolute Error (MAE), Root Mean Squared Error (RMSE), and Spearman’s rank correlation coefficient ($\rho$). Spearman’s rank correlation coefficient is defined as follows:
\vspace*{-5pt}
\begin{equation}
    \rho=1-\frac{6\sum{d_i^2}}{n(n^2-1)}
\end{equation}

$d_i$ is the difference between the ranks of each observation, and 
$n$ is the sample size. This metric assesses the correlation between the relative rankings of the actual and estimated values. MAE and RMSE metrics capture the model's precision of the individual predictions made to each sample, while Spearman’s rank correlation coefficient captures the accuracy of the relative ordering of the predictions within the test set. 

Apart from the training and evaluation of feature fusion models and unimodal models on the severity estimation task, an ablation study was also conducted on the same task by removing the Multi-head attention module of the model to prove that the module contributes significantly to the performance of the model. For all the experiments in the ablation study the best-performing feature fusion model that takes concise articulatory representations and Wav2Vec2.0-base representations as inputs was used.

\section{\textbf{Discussion}}
\vspace*{-3pt}

Table \ref{tab:results} summarizes the performance of different unimodal, and feature fusion models on the severity estimation task. Among unimodal models, those based on articulatory features outperform self-supervised speech representation models in MAE and RMSE. However, self-supervised models excel in Spearman’s rank correlation. These results suggest that the articulatory representation-based models are highly accurate in estimating values, with only a small trade-off in capturing the precise order of predictions. 

If we further investigate the performance of the 2 articulatory feature-based unimodal models, the model trained using concise articulatory representations extracted using the representation learning model proposed in this study has fared better than the model trained using TV-based FVTC articulatory features across all evaluation metrics. This shows the need for concise articulatory representations.

Another significant contribution of this work is the feature fusion model that uses both articulatory and speech representations which has outperformed all unimodal models trained on articulatory and self-supervised speech representations across all metrics. These results support the hypothesis that the fusion of articulatory and speech representations leverages more information on the biomarkers that are found in speech. 

The ablation study results summarized in table \ref{tab:ablation_results} prove that the multi-head attention mechanism used in the feature fusion model helps in creating a better fusion mechanism between articulatory and self-supervised speech feature that results in a better overall performance of the model.

\begin{table}[h!]
  \caption{\centering
  \textbf{Results of the ablation study}}
  \label{tab:ablation_results}
  \centering
  \begin{tabular}{ |l |c |c|c|}
  \hline
    \textbf{Model}&  \textbf{MAE}$\downarrow$&\textbf{RMSE}$\downarrow$&\textbf{$\rho$}$\uparrow$\\
    \hline
    Feature-Fusion with MHA&6.53&8.04&0.7587\\
    \hline
    Feature-Fusion without MHA&8.33&10.53&0.6486\\
\hline
  \end{tabular}
\end{table}

In addition, our feature fusion model outperforms the previous work on schizophrenia severity prediction which uses multimodal input from audio (TV-based FVTC) and video (Facial action unit\cite{b22} (FAU-based FVTC) in both MAE and RMSE. Although there is a slight decrease in performance on Spearman’s Rank Correlation, this minor difference has little effect on the model's overall effectiveness. These results indicate that our model excels at accurate value estimation, with only a small trade-off in precise ranking alignment. 

\section{\textbf{Conclusion}}
\vspace*{-3pt}


In conclusion, this paper presents a self-supervised representation learning model that produces concise articulatory representations that perform well on schizophrenia severity estimation. We also show that the feature fusion model that integrates them with self-supervised speech representations extracted from pre-trained models outperforms unimodal baselines and multimodal models on the severity estimation task in MAE and RMSE, demonstrating its strength in accurate prediction and consistency. Although we observe a minor reduction in Spearman's Rank Correlation, indicating a slight trade-off in capturing exact ranking, the overall performance gains highlight the advantages of fusing multiple audio-based features. The performance improvement opens a new avenue to investigate whether similar performance improvement can be expected in previous works where articulatory and self-supervised speech features were used separately. In the future, we would like to explore the possibilities of utilizing both feature fusion within modality and multimodal fusion between modalities together in a model to improve the performance of more complex mental health disorder assessment tasks.


\end{document}